# On the Uniqueness of the Quasi-Moment-Method Solution to the Pathloss Model Calibration Problem

Hisham Abubakar Muhammed, Ayotunde Abimbola Ayorinde, Francis Olutunji Okewole, Michael Adedosu Adelabu and Ike Mowete

*Abstract*— Investigations in this paper focus on establishing the uniqueness properties of the Quasi-Moment-Method (QMM) solution to the problem of calibrating nominal radiowave propagation pathloss prediction models. Nominal (basic) prediction models utilized for the investigations, were first subjected to QMM calibrations with measurements from three different propagation scenarios. Then, the nominal models were recast in forms suitable for Singular Value Decomposition (SVD) calibration before being calibrated with both the SVD and QMM algorithms. The prediction performances of the calibrated models as evaluated in terms of Root Mean Square Prediction Error (RMSE), Mean Prediction Error (MPE), and Grey Relational Grade-Mean Absolute Percentage Error (GRG-MAPE) very clearly indicate that the uniqueness of QMM-calibrations of basic pathloss models is more readily observable, when the basic models are recast in forms specific to SVD calibration. In the representative case of calibration with indoor-to-outdoor measurements, RMSE values were recorded for QMM-calibrated nominal models as 5.2639dB for the ECC33 model, and 5.3218dB for the other nominal models. Corresponding metrics for the alternative (rearranged) nominal models emerged as 5.2663dB for the ECC33 model and 5.2591dB for the other models. A similar general trend featured in the GRG-MAPE metrics, which for both SVD and QMM calibrations of all the alternative models, was recorded as 0.9131, but differed slightly (between 0.9138 and 0.9196) for the QMM calibration of the nominal models. The slight differences between these metrics (due to computational round-off approximations) confirm that when the components of basic models are linearly independent, the QMM solution is unique. Planning for wireless communications network deployment may consequently select any basic model of choice for QMM-calibration, and hence, identify relative contributions to pathloss by the model's component parts.

*Index Terms*—Gray-Relational Grade, Pathloss model calibration, Quasi-Moment-Method, Singular Value Decomposition, Uniqueness.

## I. INTRODUCTION

A PARTICULARLY useful description of propagation pathloss model calibration is that given in [1] as the process of using measurement information to fine-tune nominal models towards improving the models' prediction performances. A number of pathloss model calibration routines have been reported in the literature, the more popular of them being the Least Square Error (LSE) and the Minimum Mean Square Error (MMSE) approaches described in [2] and [3], respectively. However, some of these approaches, which though derive from existing nominal models, may not, as remarked in [4], be regarded as calibration, since in the typical case, they develop models that are structurally different from the 'base' model. For example, the 'partition-based' analysis in [2] is an MMSE approach, which derived from the 'pathloss exponent' model, with n = 2. Weighted attenuation parameters are then added to the log-distance term in a formulation whose unknowns are the additional attenuation functions. A calibration algorithm, which minimizes the error between measurement data and corresponding predicted pathloss is then utilized for the determination of the unknown quantities. The LSE approach typified by the analysis in [3] on the other hand, has the reduction of a basic model (Egli and Hata in [3]) to a linear ('slope and intercept') equivalent, as its starting point. This equivalent model is then subjected to classical linear regression. A slight variation of this approach was utilized in [4], in which the 'intercept term' of the derived model was spilt into two: one representing the 'free space' and 'frequency dependent' factors of the basic Hata model; and the other, all other distance-independent constants of the basic model. Although the process adopted in [5] involved calibration, it does not address the development of pathloss prediction models. Rather, its objective concerns the estimation of a 'calibration constant', through which 'measured pathloss' can be

The authors are with the Department of Electrical & Electronics Engineering of the University of Lagos, Akoka, Yaba, Lagos Nigeria. Emails: hmuhammad@unilag.edu.ng aayorinde@unilag.edu.ng, fokewole@unila.edu.ng madelabu@unilag.edu.ng and amowete@unilag.edu.ng

determined as the difference between this constant and received power.

Approaches, which calibrate nominal prediction models in the sense of calibration as defined by [1] include the Cuckoo-search optimization algorithm presented in [6]. In this case, four parameters of a 'UFPA' model, developed for a 5.8GHz network were calibrated to specialize the model for use in a 2.6GHz network. Another is the Singular Value Decomposition (SVD) algorithm, introduced by [7] in 2017. Essentially, the method minimizes mean square error in a calibration process similar to the SVD regression described in [8], and involving the singular value decomposition of a 'design matrix', whose entries are the parameters of the nominal model to be subjected to calibration. Unlike the algorithms of [6] and [7], the Quasi-Moment-Method (QMM) recently introduced by [9], has been shown to be able to calibrate all existing nominal models for the prediction of pathloss in both indoor and outdoor environments. One important property ascribed to the QMM by [9] is that when the 'basis' functions are linearly independent, the QMM solution is unique. However, there has been no investigation in the literature, on the veracity of this claim.

It is consequently the main objective of this paper to demonstrate that the QMM solution to the pathloss model calibration problem is indeed unique, if the basis functions are linearly independent. Four nominal models, for which linear independence has been established in [9] are selected as candidates for the investigations. In addition to QMM calibration, alternative versions of the candidate models were also subjected to SVD calibration, first, as a means of showing that SVD, under certain conditions, represents a special case of QMM: and second towards highlighting the fact that such differences as may exist in the RMSE and GRG-MAPE metrics for models, owe entirely to computational round-off approximations.

The paper, in section II, briefly presents the theoretical backgrounds for the QMM and SVD algorithms, and in section III, discusses the outcomes of the calibrations of the candidate nominal models, with measurements available from the literature. Discussions in that section also highlight the uniqueness properties of the QMM solution. Important conclusions arising from the findings of the paper are presented in section IV, which is the concluding section.

## II. THEORETICAL BACKGROUND

### A. The Quasi-Moment-Method

Let the generic nominal pathloss prediction model be described by

$$P_m = \varphi_1 + \varphi_2 + \ldots + \varphi_N \quad (1)$$

in which $\{\varphi_k\}_{k=1}^N$ may, in general, be functions of separation of transmitter and receiver (d), frequency (f), transmitter antenna height ($h_{te}$), receiver antenna height ($h_{re}$), constants, or some combinations of them. The QMM algorithm determines a set $\{\alpha_k\}_{k=1}^N$ of 'N' coefficients, such that at every measurement point $d_k$,

$$P_q(d_k) = \alpha_1 \varphi_1(d_k) + \ldots + \alpha_N \varphi_N(d_k) \cong P_{mea}(d_k), \quad (2)$$

provided that $P_{mea}(d_k)$ represents measured pathloss at $d_k$, and $P_q(d_k)$ is the corresponding pathloss predicted by the QMM-calibrated model. The solution to the problem is defined by the condition [9], [10], that the Euclidean semi-norm of the difference between measurement and corresponding prediction assumes its minimum possible value; that is

$$\varepsilon = \|P_{mea} - P_k\|_2, \quad (3)$$

is a minimum. In order to determine the unknown coefficients in Eq. (2), the Galerkin approach [11] is adopted, such that after defining 'testing' functions as identical to the 'basis' functions $\{\varphi_l\}_{l=1}^N$, and when the inner product of each $\varphi_l$ is then taken with both sides of the equation, the following matrix equation results:

$$\begin{bmatrix} \langle \varphi_1, \varphi_1 \rangle & \langle \varphi_1, \varphi_2 \rangle & \cdots & \langle \varphi_1, \varphi_N \rangle \\ \langle \varphi_2, \varphi_1 \rangle & \langle \varphi_2, \varphi_2 \rangle & \cdots & \langle \varphi_2, \varphi_N \rangle \\ \cdots & \cdots & \cdots & \cdots \\ \langle \varphi_N, \varphi_1 \rangle & \langle \varphi_N, \varphi_2 \rangle & \cdots & \langle \varphi_N, \varphi_N \rangle \end{bmatrix} \begin{bmatrix} \alpha_1 \\ \alpha_2 \\ \cdots \\ \alpha_N \end{bmatrix} = \begin{pmatrix} \langle \varphi_1, P_{mea} \rangle \\ \langle \varphi_2, P_{mea} \rangle \\ \cdots \\ \langle \varphi_N, P_{mea} \rangle \end{pmatrix}$$

(4)

or, in a more compact form,

$$[\Phi](A) = (P). \quad (4a)$$

Inner product is defined, [9], for any two functions $f_m, f_n$, by

$$\langle f_m(d_k), f_n(d_k)\rangle = \sum_{k=1}^{M} f_m(d_k) f_n(d_k). \quad (4b)$$

The desired unknown coefficients (entries to the vector $(A)$) are obtained from (4a) through the simple matrix processes of inversion $([\bullet]^{-1})$ and multiplication as

$$(A) = [\Phi]^{-1}(P). \quad (5)$$

It is the similarity of the algorithm to the method of moments utilized for the solution of electromagnetic field problems [11] that led to the name, QMM.

*B. Singular Value Decomposition Calibration*

SVD pathloss model calibration introduced by [7] may be generalized through a specification of the nominal model to be calibrated according to

$$P_{ma} = 1 + \varphi_2 + \varphi_3 + \ldots + \varphi_N, \quad (6)$$

which is similar to $P_m$ of (1), with the exception that the leading 'basis' function is now, by definition, [7], [8], set equal to unity. A 'design matrix' is then prescribed from the component parts of (6) according to

$$[D] = \begin{bmatrix} 1 & \varphi_2(d_1) & \ldots & \varphi_N(d_1) \\ 1 & \varphi_2(d_2) & \ldots & \varphi_N(d_2) \\ \ldots & \ldots & \ldots & \ldots \\ 1 & \varphi_2(d_M) & \ldots & \varphi_N(d_M) \end{bmatrix}, \quad (7)$$

and the unknown calibration coefficients are then given by [7]

$$(A) = \left[[D]^T[D]\right]^{-1}[D]^T(P_{mea}), \quad (8)$$

with $[\bullet]^T$ denoting matrix transposition. It is of interest to observe that (8) has a structure similar to those given as (9) in the MMSE approach described in [2], and as (13) by the LSE algorithm of [4]. The matrices and vectors in [2] and [4] are quite different from those in [7], so that the algorithms they represent may not be regarded as equivalents of the SVD-calibration algorithm. On the other hand, it is easy to establish that $\left[[D]^T[D]\right]$ of (8) is the same as $[\Phi]$ in (5), and that $[D]^T(P_{mea})$ of (8) is identical to $(P)$ of (5). Hence, we may conclude that the SVD model calibration algorithm represents a special case of the more generally applicable QMM.

The computational results presented in this paper derive from the QMM and SVD calibrations of some nominal models and their associated alternatives, using field measurement data available, through the use of the commercial graph digitizer 'GETDATA', from [5] and [6]; and raw data provided by [12].

*C. Nominal Models: QMM Calibration.*

The nominal models subjected to QMM calibration using measurements from [6] include the ECC33 model defined by

$$\{\varphi_{ECC}\} = \begin{vmatrix} 92.4, 20\log_{10}d, 20\log_{10}f, \\ 20.41, 9.83\log_{10}d, \\ \log_{10}f(7.894 + 9.56\log_{10}f), \\ -13.98\log_{10}(h_{te}/200), \\ \log_{10}(h_{te}/200)(-5.8(\log 10 d)^2), \\ -42.57(\log_{10}h_{re} - 0.585), \\ 13.7\log_{10}f(\log_{10}h_{re} - 0.585) \end{vmatrix}, \quad (9)$$

and the Stanford University Interim (SUI) model for which

$$\{\varphi_{SUI}\} = \begin{cases} 20.7412, 52.15\log_{10}(d/100), \\ 6\log_{10}(f/2000), \\ -10.8\log_{10}(h_{re}/2000), 8.9 \end{cases}. \quad (10)$$

Calibration with data from [6] also involved the UFPA model defined here by

$$\{\varphi_{UFPA}\} = \begin{cases} 1, -2.4((h_{te} + h_{re})/6.2)(30/f_{GHz}), \\ 20\log_{10}d, 12\log_{10}f_{MHz} \end{cases}, \quad (11)$$

as well as the nominal Ericsson and Lee models defined, respectively, by

$$\{\varphi_{Eric}\} = \begin{cases} 36.2, 30.2\log_{10}d, -12\log_{10}h_{te}, \\ 0.1\log_{10}h_{te}\log_{10}d, -3.2(\log_{10}11.75h_{re})^2, \\ \log_{10}f(44.49 - 4.78\log_{10}f) \end{cases} \quad (12)$$

and

$$\{\varphi_{Lee}\} = \begin{cases} 124, 30.5\log_{10}(d/1.6), 30\log_{10}(f/900) \\ -3.001 \end{cases}. \quad (13)$$

The nominal models defined by (9), (10), (12), and (13) were also QMM-calibrated, with measurement data from [12]. For calibration with 26GHz indoor measurements in [5], the nominal models of (9) and (10), as well as the WINNER-II and ITU-R (LOS) models were considered. These latter models are defined by

$$\{\varphi_{WIN}\} = \{46.8, 18.7\log_{10}d, 20\log_{10}(f/5.0)\}, \quad (14)$$

and

$$\{\varphi_{ITU}\} = \{6.0, 25\log_{10}d, 20\log_{10}f, -28\}. \quad (15)$$

It should be noted that for the nominal SUI model in this case, the first two terms of (10) modify to $100.7412$ and $129.8875\log_{10}(d/100)$, respectively.

### D. Alternative Models: QMM and SVD Calibration

Calibration with the SVD algorithm requires [7], [8] that the nominal models be expressed in the form of (6). Accordingly, alternatives to the models of (9) to (15) were specified as

$$\{\varphi_{ECC-A}\} = \begin{vmatrix} 1,(91.4 + 20\log_{10}d + 20\log_{10}f + 0.09(k-1)), \\ (20.41 + 9.83\log_{10}d \\ +\log_{10}f(7.894 + 9.56\log_{10}f) + 0.012(k-1) \\ (-13.98\log_{10}(h_{te}/200) + \\ \log_{10}(h_{te}/200)(-5.8(\log_{10}d)^2) \\ -42.57(\log_{10}h_{re}-0.585) + \\ 13.7\log_{10}f(\log_{10}h_{re}-0.585) + 0.031(k-1) \end{vmatrix}, \quad (16)$$

with

$$\{\varphi_{SUI-A}\} = \begin{cases} 1,(19.7412 + 52.15\log_{10}(d/100)), \\ (6\log_{10}(f/2000) + 5.22(k-1)), \\ 8.9 - 10.8\log_{10}(h_{re}/2000) + 0.95(k-1) \end{cases}, \quad (17)$$

whilst

$$\{\varphi_{UFPA-A}\} = \begin{cases} 1,(-2.4((h_{te}+h_{re})/6.2)(30/f_{GHz})) \\ +0.58(k-1), \\ (20\log_{10}+12\log_{10}f_{MHz}) \end{cases}, \quad (18)$$

and

$$\{\varphi_{Eric-A}\} = \begin{cases} 1,(35.2 + 30.2\log_{10}d, -12\log_{10}h_{te}), \\ (0.1\log_{10}h_{te}\log_{10}d, -3.2(\log_{10}11.75h_{re})^2 \\ +0.58(k01)), \\ (\log_{10}f(44.49 - 4.78\log_{10}f)) \end{cases}. \quad (19)$$

Also,

$$\{\varphi_{Lee-A}\} = \begin{cases} 1,(123 + 30.5\log_{10}(d/1.6)), \\ (30\log_{10}(f/900) - 3.001 + 2.85(k-1)) \end{cases}, \quad (20)$$

with

$$\{\varphi_{WIN-A}\} = \begin{cases} 1,(45.8 + 18.7\log_{10}d), \\ (20\log_{10}(f/5.0) + 0.065(k-1)) \end{cases}, \quad (21)$$

and

$$\{\varphi_{ITU-A}\} = \begin{cases} 1,(5 + 25\log_{10}d), \\ 20\log_{10}f - 28.0 + 0.095(k=1) \end{cases}. \quad (22)$$

The terms generically represented by $constant*(k-1)$, $k=1,2,\ldots,M$, in each of (16) to (22) were introduced in order to remove the 'ill-condition' character assigned by MATLAB to the matrix $[[D]^T[D]]$ in the course of implementing the SVD calibration algorithm.

## III. COMPUTATIONAL RESULTS AND DISCUSSIONS

For the nominal models described by (9) to (15) (as may apply) the calibration coefficients obtained are presented below, as follows.

### A. Calibration Coefficients: Nominal Models

#### 1) **Calibration With Data from Route 2 of [6]**

Outcomes of the QMM-calibration of the basic models are defined by the calibration coefficients given below as

$$\{\alpha_{ECC}\} = \begin{cases} 0.8486 & 2.5392 & 0.8032 & 1.1887 \\ 6.2557 & 11.5896 & -4.0637 & 26.4503 \\ -0.3678 & 16.8936 & & \end{cases}, \quad (23a)$$

for the ECC33 model. Corresponding results for the SUI, UFPA, Ericsson, and Lee models emerged as

$$\{\alpha_{SUI}\} = \begin{cases} 1.5301 & 0.2598 & 40.0387 \\ 2.4291 & -1.8999 & \end{cases}, \quad (23b)$$

$$\{\alpha_{UFPA}\} = \begin{cases} 33.7745 & -0.1378 & 0.6775 \\ 0.9023 & & \end{cases}, \quad (23c)$$

$$\{\alpha_{Eric}\} = \begin{cases} 1.0014 & 0.1003 & -1.7322 \\ 65.6489 & -1.2733 & 0.2080 \end{cases}, \quad (23d)$$

and

$$\{\alpha_{Lee}\} = \begin{cases} 0.1042 & 0.4439 & -1.4623 \\ -35.4671 & & \end{cases}, \quad (23e)$$

respectively.

#### 2) **Calibration With Data for Route A of [12]**

In this case, the model calibration coefficients were obtained as

$$\{\alpha_{ECC}\} = \begin{Bmatrix} 1.1485 & -1.9748 & 17.5363 & -0.5925 \\ -6.6973 & 12.9777 & 2.7195 & 4.2358 \\ 3.6562 & -64.9787 & & \end{Bmatrix},$$

(24a)

for the ECC33 model; and for the SUI, Ericsson, and Lee models, as

$$\{\alpha_{SUI}\} = \begin{Bmatrix} 0.7587 & 0.3735 & -552.5412 \\ -1.0549 & -1.5267 & \end{Bmatrix},$$

(24b)

$$\{\alpha_{Eric}\} = \begin{Bmatrix} 1.6632 & 0.2809 & 0.1330 \\ 88.6193 & -2.2254 & 0.8839 \end{Bmatrix},$$

(24c)

and

$$\{\alpha_{Lee}\} = \begin{Bmatrix} 0.3979 & 0.6431 & -0.3840 \\ -12.3772 & & \end{Bmatrix},$$

(24d)

respectively.

**3) Calibration With Data from Fig.3a of [5]**

In this case, the calibration coefficients were obtained as

$$\{\alpha_{ECC}\} = \begin{Bmatrix} -0.0713 & 1.0511 & 1.0514 & 0.3159 \\ 1.8884 & 0.3530 & 0.4458 & -0.8269 \\ 0.0317 & -1.0002 & & \end{Bmatrix}.$$

(25a)

for the ECC33 model, and for the SUI, ITU-R, and WINNERII models, as

$$\{\alpha_{SUI}\} = \begin{Bmatrix} 0.3639 & 0.1052 & 7.2696 \\ 0.4770 & -3.9531 & \end{Bmatrix},$$

(25b)

$$\{\alpha_{WIN}\} = \{4.1277 \quad 0.7307 \quad -9.1590\},$$

(25c)

and

$$\{\alpha_{ITU}\} = \{0.5110 \quad 0.6121 \quad 0.2297 \quad -1.3013\},$$ (25d)

respectively.

*B. Calibration Coefficients: Alternative Models*

The alternative models of (16) to (22) were subjected to both QMM and SVD calibration with the same sets of measurement data. Calibration coefficients obtained in each case are presented below. In each case, the rows labelled "Q" identify the coefficients associated with QMM calibration, and those labelled "S", SVD calibration.

**1) Calibration With Data from Route 2 of [6]**

For this case, model calibration coefficients for the ECC33 models were computed as

$$\{\alpha_{ECC-A}\} = \begin{Bmatrix} Q: & 215.6466 & -1.2728 & -0.9728 & 1.3469 \\ S: & 454.6644 & -4.5185 & 4.4421 & -0.6746 \end{Bmatrix}.$$

(26a)

Corresponding results were obtained for the SUI models as

$$\{\alpha_{SUI-A}\} = \begin{Bmatrix} Q: & 156.3261 & -0.8235 & 0.0449 \\ S: & 156.3596 & -0.8245 & 0.0449 \end{Bmatrix},$$

(26b)

and for the UFPA models, as

$$\{\alpha_{UFP-A}\} = \begin{Bmatrix} Q: & 403.6123 & 0.4762 & -2.1386 \\ S: & 404.9230 & 0.4780 & -2.1498 \end{Bmatrix}.$$

(26c)

Calibration coefficients for the Ericsson and Lee models were obtained in this case, as

$$\{\alpha_{Eric-A}\} = \begin{Bmatrix} Q: & 638.5167 & -1.3739 & -26.7029 \\ S: & 653.2062 & -1.4162 & -27.4656 \end{Bmatrix},$$

(26d)

and

$$\{\alpha_{Lee-A}\} = \begin{Bmatrix} Q: & 386.9501 & -1.3851 & 0.0961 \\ S: & 391.5694 & -1.4097 & 0.0973 \end{Bmatrix},$$

(26e)

respectively.

**2) Calibration With Measurement Data for Route A of [12]**

With the use of measurements available from [12], the model calibration coefficients for the alternative ECC33 model were obtained as

$$\{\alpha_{ECC-A}\} = \begin{Bmatrix} Q: & -211.3346 & 0.0761 & 17.7241 & -7.6040 \\ S: & -383.1484 & 0.0866 & 26.6139 & -11.2481 \end{Bmatrix},$$

(27a)

whilst those for the SUI model emerged as

$$\{\alpha_{SUI-A}\} = \begin{Bmatrix} Q: & 121.6438 & -0.1378 & 0.0093 \\ S: & 102.6361 & -0.1373 & 0.4576 \end{Bmatrix}.$$

(27b)

Corresponding respective results for the Ericsson and Lee models are

$$\{\alpha_{Eric-A}\} = \begin{Bmatrix} Q: & 198.3418 & -0.2375 & -4.8761 \\ S: & 198.2258 & -0.2371 & -4.8696 \end{Bmatrix},$$

(27c)

and

$$\{\alpha_{Lee-A}\} = \begin{Bmatrix} Q: & 161.2499 & -0.2365 & 0.0201 \\ S: & 161.1078 & -0.2358 & 0.0201 \end{Bmatrix}.$$

(27d)

**3) Calibration With Measurement Data from Fig. 3a of [5]**

For the four nominal models calibrated with measurement data available from [5], the model calibration coefficients were obtained as

$$\{\alpha_{ECC-A}\} = \begin{cases} Q: & -52.5628 \quad -3.3234 \quad 8.7702 \quad 0.9793 \\ S: & -78.8487 \quad -1.6592 \quad 6.4088 \quad 0.0036 \end{cases},$$
(28a)

in the case of the nominal ECC33 model, and

$$\{\alpha_{SUI-A}\} = \begin{cases} Q: & 86.2491 \quad 0.2299 \quad -0.0385 \\ S: & 86.2481 \quad 0.2299 \quad -0.0385 \end{cases},$$
(28b)

for the nominal SUI model. Calibration coefficients were obtained as

$$\{\alpha_{WIN-A}\} = \begin{cases} Q: & -9.2209 \quad 1.5910 \quad -1.1009 \\ S: & -9.4853 \quad 1.5969 \quad -1.1081 \end{cases},$$
(28c)

for the nominal WINNERII model, and as

$$\{\alpha_{ITU-A}\} = \begin{cases} Q: & 87.5782 \quad 1.1956 \quad -0.7595 \\ S: & 87.5298 \quad 1.1945 \quad -0.7582 \end{cases},$$
(28d)

for the nominal ITU-R model.

### C. Evaluation of Pathloss Predictions

Pathloss predicted by the calibrated nominal models (and their alternatives) defined by (9) to (28) are evaluated in this section, first through graphical comparisons with the measurements from which they derive; and then, using the performance metrics of MPE, RMSE, and GRG-MAPE.

The pathloss profiles displayed Fig. 1 compare predictions by the calibrated ECC33, UFPA, SUI, Ericsson, and Lee models with measurement data available from Fig. 2 ('measurement route 2') of [6]. Two things are immediately observable from the curves; first, the profiles of pathloss predicted by the QMM-calibrated nominal models, in all cases, differ significantly from those of the corresponding QMM-calibrated alternative models. And second, profiles of the SVD- and QMM-calibrated alternative models are virtually identical, also in all cases.

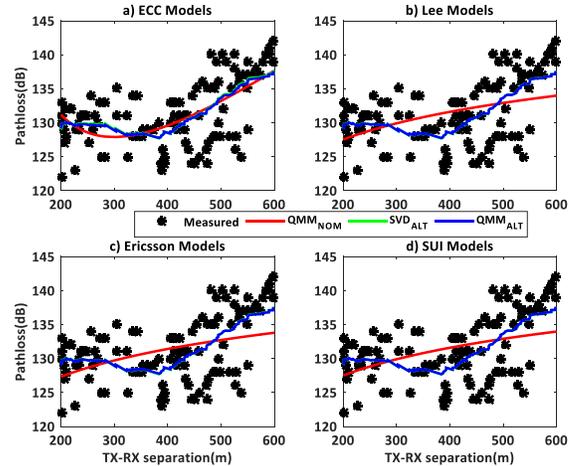

**Fig. 1.** Comparison of pathloss predicted by calibrated models with corresponding measurements from [6].

This latter observation is underscored by the prediction profiles of Fig. 2, which essentially compare pathloss predicted by the SVD- and QMM-calibrated alternative models as defined by the sets of (16) to (22), and (26a) to (26e).

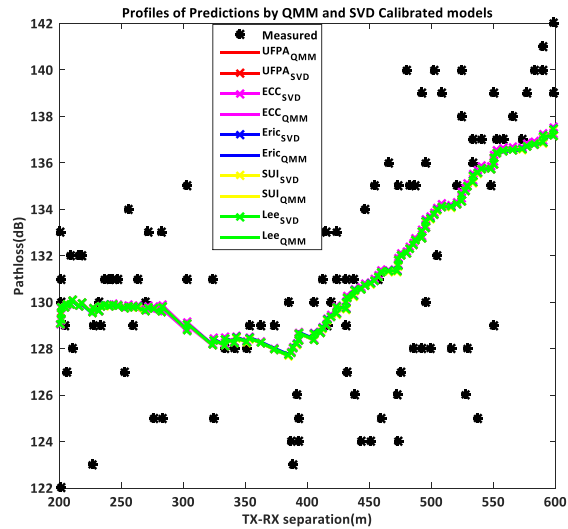

**Fig. 2.** Profiles of pathloss predicted by alternative models SVD- and QMM-calibrated using measurements from [6].

MPE and RMSE metrics recorded by the calibrated models are displayed in Table I.

TABLE I
PERFORMANCE METRICS FOR PROFILES OF FIG. 1

| Model | MPE(dB) | | | RMSE(dB) | | |
|---|---|---|---|---|---|---|
| | $Q_{NOM}$ | $Q_{ALT}$ | SVD | $Q_{NOM}$ | $Q_{ALT}$ | SVD |
| ECC | -0.0021 | 0.0189 | -0.0594 | 3.9836 | 3.8909 | 3.8909 |
| Ericsson | 0.0034 | -0.0013 | 0.0153 | 4.3942 | 3.8904 | 3.8904 |
| SUI | 0.1832 | 0.0028 | 0.0026 | 4.3980 | 3.8906 | 3.8904 |
| Lee | -0.0060 | 0.0006 | 0.0014 | 4.3942 | 3.8905 | 3.8904 |
| UFPA | -0.0069 | -0.0059 | 0.0012 | 4.3942 | 3.8909 | 3.8909 |

MPE metrics for all the calibrated models are generally impressive, ranging between 0.006dB for the QMM-calibrated alternative model and 0.1832dB for the QMM-calibrated nominal Ericsson model. On the other hand, RMSE metrics are virtually the same at about 3.89dB for all the alternative models calibrated with the QMM and SVD algorithms. RMSE metrics for the nominal models calibrated with the QMM algorithm are generally slightly higher (about 0.09dB for ECC33 and about 0.5dB for the other four models) than those for the corresponding alternative models. These differences may be attributed to the computational round-off approximations involved in the implementation of the algorithms.

Profiles of the pathloss predicted by the nominal and alternative models calibrated with measurement data provided by [12] are displayed in Figs. 3 and 4.

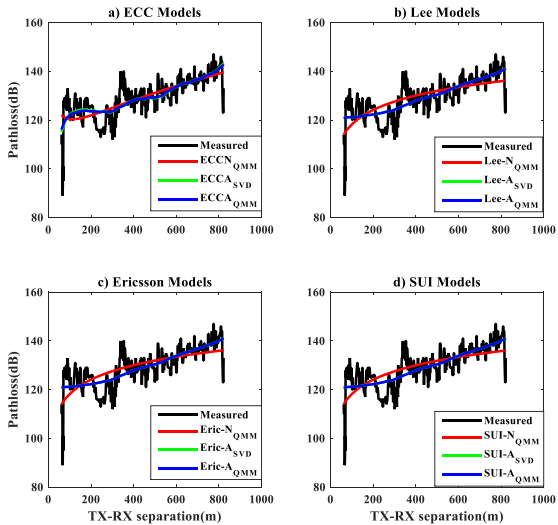

**Fig. 3.** Comparison of pathloss predicted by calibrated models with corresponding measurements from [12].

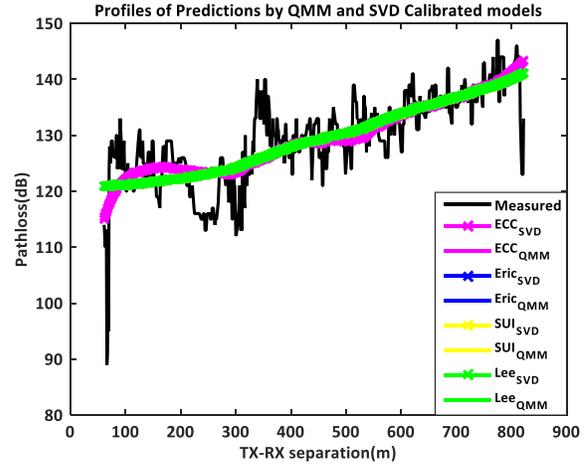

**Fig. 4.** Profiles of pathloss predicted by alternative models SVD- and QMM-calibrated using measurements from [12].

The profiles of Fig. 3, with the exception of Fig. 3a, share the features described concerning Fig. 1, for calibrated nominal and corresponding alternative models. This exception is that the SVD and QMM calibrations of the alternative ECC33 model by measurement data from [12-route A], do not yield identical prediction outcomes; especially in regions relatively close to the transmitter. The exception clearly manifests in Fig. 4, whose features are otherwise, the same as those described for the profiles of Fig. 2.

TABLE II
PERFORMANCE METRICS FOR PROFILES OF FIG. 3

| Model | MPE(dB) | | | RMSE(dB) | | |
|---|---|---|---|---|---|---|
| | $Q_{NOM}$ | $Q_{ALT}$ | SVD | $Q_{NOM}$ | $Q_{ALT}$ | SVD |
| ECC | -0.0031 | 0.0130 | -0.0055 | 5.5791 | 5.3024 | 5.2970 |
| SUI | 0.0747 | -0.0037 | 0.0038 | 6.0389 | 5.4413 | 5.4413 |
| Ericsson | 0.0047 | -0.0070 | -0.0052 | 6.0383 | 5.4413 | 5.4413 |
| Lee | 0.0045 | -0.0056 | -0.0001 | 6.0383 | 5.4413 | 5.4413 |

MPE and RMSE metrics for the calibrated models as displayed in Table II also follow the trend earlier described for the metrics of Table I. In this case, the MPE metrics, like those recorded for the profiles of Fig. 1 are also excellent, ranging between -0.0001dB for SVD calibrated alternative Lee model and 0.0747dB, for the QMM-calibrated nominal SUI model. The RMSE metrics of Table II reveal that despite the differences in the profiles of Fig. 4, the difference in RMSE between the SVD- and QMM-calibrated alternative ECC33 model is 0.0234dB. It is also readily observed from Table II that RMSE is exactly the same for the Ericsson, SUI, and Lee models, although RMSE recorded for the calibrated nominal models is slightly higher (0.2767dB) than those for the calibrated alternative models. For both calibrated nominal and alternative models, RMSE metrics recorded for the ECC33 cases are slightly smaller (about 0.46dB for the nominal models and 0.12dB for the alternative models) than the common metric for the other three.

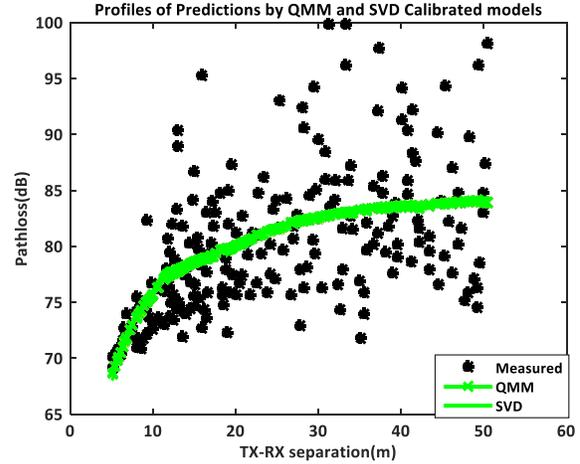

Fig. 6. Profiles of pathloss predicted by alternative models SVD- and QMM-calibrated using measurements from [5].

The profiles of Fig. 6, like those of Fig. 2, reveal that QMM and SVD calibration of the alternative models lead to identical predictions by all the calibrated models.

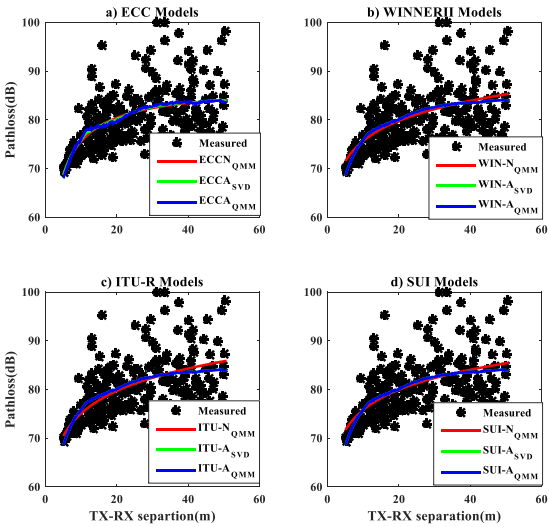

Fig. 5. Comparison of pathloss predicted by calibrated models with corresponding measurements from [5].

TABLE III
PERFORMANCE METRICS FOR THE PROFILES OF FIG. 5

| Model | MPE(dB) | | | RMSE(dB) | | |
|---|---|---|---|---|---|---|
| | $Q_{NOM}$ | $Q_{ALT}$ | SVD | $Q_{NOM}$ | $Q_{ALT}$ | SVD |
| ECC | 0.0766 | 0.0020 | -0.0082 | 5.2639 | 5.2663 | 5.2591 |
| SUI | -0.1414 | -0.0053 | -0.0043 | 5.3128 | 5.2591 | 5.2591 |
| WIN II | -0.0021 | -0.0008 | -0.0027 | 5.3110 | 5.2591 | 5.2591 |
| ITU | 0.0490 | -0.0013 | -0.0017 | 5.3282 | 5.2591 | 5.2591 |

A third example considered in this paper concerns the calibration of basic and alternative models, using measurements for indoor millimeter wave scenario associated with Fig. 3a of [5]. The candidate models in this case are the ECC33, SUI, WINNER II, and ITU-R models, for which prediction performance evaluation through comparisons with measurement is provided by the curves of Fig. 5. The profiles follow the same general pattern as those of Figs. 1 and 3, in that where the SVD- and QMM-calibrated alternative models provide virtually identical predictions, those for the QMM-calibrated nominal models are slightly different, in all cases.

According to the metrics of Table III, MPE metrics recorded for the calibrated models ranged between 0.0008dB for the QMM-calibrated alternative WINNER model to 0.1414dB for the QMM-calibrated nominal SUI model. Again, with the exception of the QMM-calibrated alternative ECC33 model, all the calibrated alternative models (including the SVD calibrated alternative ECC33 model) recorded RMSE values of 5.2951dB, which differed from that of the QMM-calibrated ECC33 model by only 0.0072dB. The biggest difference of 0.0691dB in RMSE between a QMM-calibrated nominal model and its corresponding alternative model was in this case, recorded by ITU case.

The virtually identical RMSE values recorded by all the calibrated models of the three examples considered in this paper very clearly support the uniqueness property of the QMM pathloss model calibration algorithm as defined in [9], [15]. It has however been suggested, [6], [13], [14], that a better assessment of pathloss model prediction performance is offered by the Grey Relational Grade Mean Absolute Percentage Error, GRG-MAPE. Because [13] and [14] presented the generic GRG-MAPE algorithm without specializing it to the case of pathloss prediction, it is helpful to provide details of the algorithm's use for the results presented here.

*D. The Grey Relational Grade MAPE Algorithm*

Suppose that $P_{mea}(d_k), k = 1, 2, \ldots, M$ represents field measurement data with which the generic nominal (or alternative) model is calibrated to yield the pathloss prediction function denoted by $P_{pre}(d_k)$. And let $[(P_{mea})_{max}, (P_{mea})_{min}]$ represent the maximum and minimum values of $P_{mea}(d_k)$, with $[(P_{pre})_{max}, (P_{pre})_{min}]$ denoting the corresponding quantities for $P_{pre}(d_k)$. The first step in the GRG-MAPE algorithm is that of 'normalization', for which the quantities

$$(P_{mea})_n = \frac{(P_{mea})_{max} - P_{mea}}{(P_{mea})_{max} - (P_{mea})_{min}}, \quad (29a)$$

and

$$(P_{pre})_n = \frac{(P_{pre})_{max} - P_{pre}}{(P_{pre})_{max} - (P_{pre})_{min}} \quad (29b)$$

are defined. Next, the 'deviation sequence' is determined for each of the measurement and prediction data according to

$$\Delta P = |(P_{mea})_n - (P_{pre})_n|, \quad (30)$$

from which the Grey Relational Coefficient is obtained as

$$\varsigma(k) = \frac{(\Delta P)_{min} + \xi(\Delta P)_{max}}{\Delta P + \xi(\Delta P)_{max}}. \quad (31)$$

Equation (31) expresses the relationship between measured and predicted pathloss, and the 'distinguishing' or 'identification coefficient' symbolized by '$\xi$', is prescribed as [13], [14], $\xi \in [0, 1]$; the commonly utilized value of $\xi = 0.5$ was adopted for the computational results of this paper.

The Grey Relational Grade here denoted by $\rho_{GRG}$ is given by

$$\rho_{GRG} = \frac{1}{M} \sum_{k=1}^{M} \varsigma(k) \quad (32)$$

and towards, thereafter, determining the GRG-MAPE, 'normalized' absolute prediction error is evaluated as

$$\varepsilon_a(k) = \frac{|P_{mea}(k) - P_{pre}(k)|}{P_{mea}(k)}. \quad (33)$$

And its mean is then given by

$$\bar{\varepsilon}_a = \frac{1}{M} \sum_{k=1}^{M} \varepsilon_a(k), \quad (33a)$$

so that MAPE is obtained [13], [14] as

$$\rho_{MAPE} = 1 - \bar{\varepsilon}_a. \quad (34)$$

Hence, GRG-MAPE is determined according to

$$\rho_{GRG-MAPE} = |\sigma \rho_{GRG} + \beta \rho_{MAPE}|. \quad (35)$$

In this paper, the quantities represented by $\sigma$ and $\beta$ are assigned their typically utilized values of 0.1 and 0.9, respectively, [14].

The GRG-MAPE metrics obtained for the nominal and alternative models calibrated with measurements from [6] are displayed in Table IV.

TABLE IV
GRG-MAPE METRICS FOR THE PROFILES OF FIG. 1

| Model | $QMM_{ALT}$ | $QMM_{NOM}$ | $SVD_{ALT}$ |
|---|---|---|---|
| ECC33 | 0.9416 | 0.9416 | 0.9421 |
| SUI | 0.9420 | 0.9370 | 0.9420 |
| ERICSSON | 0.9418 | 0.9369 | 0.9420 |
| LEE | 0.9419 | 0.9370 | 0.9420 |
| UFPA | 0.9420 | 0.9416 | 0.9420 |

Metrics in the table reveal that GRG-MAPE is virtually the same for all the models. In the case of QMM-calibrated alternative models for example, the difference between the smallest and largest metrics is 0.0004, and the corresponding quantities for the SVD-calibrated alternative models and QMM-calibrated nominal models are 0.0001 and 0.0047, respectively.

This general trend is also evident in Tables V and VI, which display GRG-MAPE metrics for the pathloss profiles of Figs. 3 and 5, respectively.

TABLE V
GRG-MAPE METRICS FOR THE PROFILES OF FIG. 3

| Model | QMM$_{ALT}$ | QMM$_{NOM}$ | SVD$_{ALT}$ |
|---|---|---|---|
| ECC33 | 0.9351 | 0.9368 | 0.9357 |
| SUI | 0.9339 | 0.9403 | 0.9339 |
| ERICSSON | 0.9339 | 0.9403 | 0.9339 |
| LEE | 0.9339 | 0.9403 | 0.9339 |

It is readily observed from Table V that for each of columns 2 to 4, only the metric for the ECC33 calibrated models differs from those for the other three models. These differences range from 0.0012 (QMM$_{ALT}$), through 0.0035(QMM$_{NOM}$), to 0.0018, in the case of the SVD calibration of the alternative models.

TABLE VI
GRG-MAPE METRICS FOR THE PROFILES OF FIG. 5

| Model | QMM$_{ALT}$ | QMM$_{NOM}$ | SVD$_{ALT}$ |
|---|---|---|---|
| ECC33 | 0.9120 | 0.9138 | 0.9131 |
| SUI | 0.9131 | 0.9190 | 0.9131 |
| WINNERII | 0.9131 | 0.9193 | 0.9131 |
| ITU-R | 0.9131 | 0.9196 | 0.9131 |

A slightly different pattern is displayed in Table VI. First, GRG-MAPE is exactly the same for all SVD-calibrated alternative models; and as obtained for the metrics in Table V, only the metric for the QMM-calibrated alternative ECC33 model is different in column 2: the difference in this case being 0.0011. Metrics in column 3 of the Table all slightly differ, with the largest difference (between the calibrated nominal ECC33 and ITU-R models) being 0.0058.

One possible application of the results of this paper derives from the fact that because the calibration solution is unique for all basic models that satisfy the requirements set forth in [9], QMM offers the possibility of physical interpretations that could be of important use to the network planning process. As an illustrative example, contributions to net pathloss due to the component parts of the nominal ECC33 model calibrated with measurements of route 2 of [6] are displayed in Fig. 7.

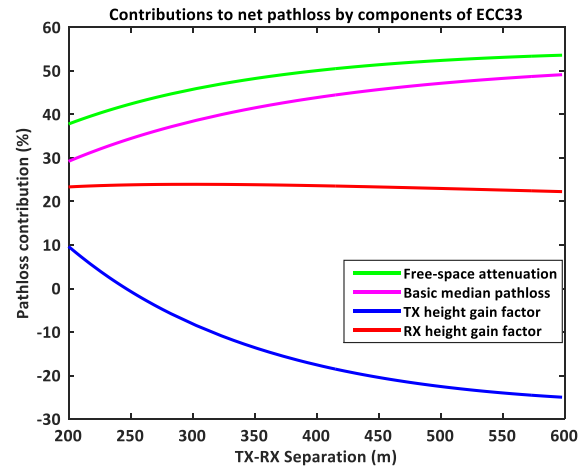

**Fig. 7.** Percentage contributions to net pathloss as predicted by components of nominal ECC33 QMM-calibrated using measurements from [6].

According to the profiles of Fig. 7, the ECC33 nominal model QMM-calibrated with measurements of route 2 of [6] predicts that percentage contributions by the 'free space attenuation factor ranges from about 37% 200m away from the transmitter, to about 53%, 600 m away. Corresponding contributions from the 'basic median pathloss' component are about 29% and 47%, respectively, whilst contributions from the 'Rx height gain factor' remained more or less constant, at about 23%. These contributions are moderated by those due to the 'Tx height gain factor', which ranged between 9% and -25%.

Thus, if for example, the factors of interest to network planning include relative contributions to net pathloss by base station (BS) antenna height, the calibration of the basic ECC33 model could offer useful information. Similar interpretations are readily available from any of the QMM/SVD calibrated models.

## IV. CONCLUSION

This paper has systematically investigated the uniqueness properties of the Quasi-Moment-Method (QMM) approach to the calibration of basic (nominal) pathloss models. Measurements available for three different propagation scenarios (including a 2.6GHz LTE network for a city-forest environment, a GSM 1800 network in a university campus set up, and a 26GHz network for indoor communications) were utilized for the calibration of nominal and alternative pathloss prediction models. The prediction performances of the calibrated models were evaluated with the use of the common performance metrics of Mean Prediction Error (MPE), and for the purposes of assessing the uniqueness of solution, Root Mean Square Error (RMSE) and Grey Relational Grade-Mean Absolute Percentage Error (GRG-MAPE). The MPE metrics were in all cases, as impressive as those reported elsewhere, [9], [15], and the RMSE and GRG-MAPE metrics very clearly indicate that the QMM solution to the pathloss model calibration problem is indeed unique, when the conditions specified in [9] and [15] are satisfied. For example, the difference between the largest and smallest recorded RMSE values for the models calibrated with measurements from the 2.6GHz network (Table I) is 0.5038dB, with 0.7415dB and 0.0331 as the corresponding values for the 1800MHz (Table II) and 26GHz (Table III), respectively. In the case of GRG-MAPE, these differences emerged as 0.0005 for the 2.6GHz network (Table IV) and 0.0066 for the 1800MHz network (Table V), 0.0076 for the 26GHz network, as can be seen from Table VI. These absolutely very small differences clearly owe to computational round-off approximations involved in the implementation of the algorithms.

Two other important conclusions arising from the results presented in the paper are first, that when the nominal models are adjusted to forms suitable for SVD calibration (shown here to be a special case of QMM calibration), the influence of computational round-off approximations on both RMSE and GRG-MAPE is considerably reduced. Second, by disaggregating net pathloss into contributions due to components of the calibrated model, it becomes possible to separately quantify the influence of the parameters of the operational environment on net pathloss.

REFERENCES

[1] J. Zhang, C. Gentile, and W. Garey, "On the Cross-Application of Calibrated Pathloss Models Using Area Features: Finding a way to determine similarity between areas," *IEEE Antennas and Propagation Magazine,* vol. 62, No. 1, Feb. 2020. pp. 40-50, doi: 10.1109/MAP.2019.2943272

[2] G. D. Durgin, T. S. Rappaport, and H. Xu, "Partition-based path loss analysis for in-home and residential areas at 5.85 GHz," in *IEEE GLOBECOM 1998* (Cat. NO. 98CH36250), 1998, vol. 2, pp. 904-909, 1998, doi: 10.1109/GLOCOM.1998.776862.

[3] T. Jawhly, and R. C Tiwari, "The special case of Egli and Hata model optimization using least-square approximation method", *SN Applied Sciences,* vol. 2, pp. 1-10, 2020, doi: 10.1007/s42452-020-3061-0

[4] K. Diawuo, K. A. Dotche, and T. Cumberbatch, "Data Fitting to Propagation Model Using Least Square Algorithm: A Case Study in Ghana," *International Journal of Engineering Sciences,* vol. 2, no. 6, pp. 226-230, June 2013.

[5] D. Pimiemta-del-Valle, L. Mendo, J. M. Riera, and P. Garcia-del-Pino, "Indoor LOS Measurements and Modeling at 26, 32, and 39 GHz Millimeter-wave Frequency Bands," *Electronics,* vol. 9, no. 11, pp. 1867, 2020, doi: 10.3390/electronics9111867

[6] A. Carvalho, I. Batalha, M. Alcantara, B. Castro, F. Barros, J. Araujo, & G. Cavalcante, "Empirical Path Loss Model in City-forest Environment for Mobile Communications", *Journal of Communication and Information Systems,* vol. 36(1), pp. 70-74, 2021, doi: 10.14209/jcis.2021.7

[7] B. Allen, S. Mahato, Y. Gao, and S. Salous, "Indoor-to-outdoor empirical path loss modelling for femtocell networks at 0.9, 2, 2.5 and 3.5 GHz using singular value decomposition," *IET microwaves antennas & propagation,* 2017, vol. 11(9), pp. 1203-1211, 2017, doi: 10.1049/iet-map.2016.0416

[8] J. Mandel, "Use of the Singular Value Decomposition in Regression Analysis", *The American Statistician,* vol. 36, no. 1, pp. 15-24, 1982, doi: 10.1080/00031305.1982.10482771

[9] S. A. Adekola, A. A, Ayorinde, F. O. Okewole, and I. Mowete, "A Quasi-Moment-Method empirical modelling for pathloss prediction," *International Journal of Electronics Letters*, pp. 1-14, 2021, doi: 10.1080/21681724.2021.1908607

[10] G. Dahlquist, and A. Björck, (Translated by Ned Anderson), *Numerical Methods*, Mineola, New York, USA: Dover Publications Inc., Section 4.2, pp. 88-92, 1974

[11] R. F. Harrington, "Matrix Methods for Field Problems," *In Proceedings of the IEEE,* vol. 55, no. 2, pp. 136-149, 1967, doi: 10.1109/PROC.1967.5433

[12] S. I. Popoola, A. A. Atayero, and O. A. Popoola, "Comparative assessment of data obtained using empirical models for path loss predictions in a university campus environment", *Data in Brief*, vol. 18, pp. 380-393, 2018, doi: 10.1016/j.dib.2018.03.040

[13] Y. Shui, F. Li, J. Yu, W. Chen, C. Li, K. Yang, and F. Chang, "Vehicle-to-Vehicle Radio Channel Characteristics for Congestion Scenario in Dense Urban Region at 5.9GHz," *International Journal of Antennas and Propagation,* vol. 2018, Article ID 1751869, doi: 10.1155/2018/1751869

[14] J. Yu, C. Li, K. Yang and W. Chen, "GRG-MAPE and PCC-MAPE Based on Uncertainty-Mathematical Theory for Path-Loss Model Selection," presented at 2016 *IEEE 83rd Vehicular Technology Conference* (VTC Spring), 2016, pp. 1-5, 2016, doi: 10.1109/VTCSpring.2016.7504265.

[15] A. A. Ayorinde, H. A. Muhammed, F. O. Okewole, and A. I. Mowete, "A novel propagation pathloss model calibration tool," *International Journal on Communication Antennas and Propagation (IRECAP)*, vol. 11, no. 3, pp. 166-180, 2021, doi: 10.15866/irecap.v113.19796

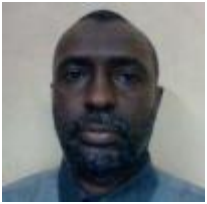
**Hisham Muhammed** received the B. Eng. and M. Sc. degrees in Electrical Engineering, from the University of Maiduguri (Nigeria) in 1991 and the University of Lagos, in 1998, respectively. He has been a lecturer at UNILAG since 2002. His research interests extend over antennas and propagation, Biomedical Instrumentation, and software development. Muhammed's draft Ph.D. thesis on "A matrix vector potential analysis of the bi-elliptical toroidal helical antenna", is at about ready for oral examination.

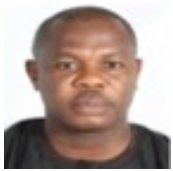
**Ayotunde Ayorinde** received the B.Sc., M. Sc. and Ph.D. degrees in Electrical Engineering, from the University of Lagos, (UNILAG) Nigeria, in 1990, 1993, and 2009, respectively. His current research interests include electromagnetic fields, microwave engineering, antennas and propagation. Dr. Ayorinde, a Nigerian Registered Engineer, is a Senior Lecturer at the University of Lagos, and is also a member of the British Institution of Engineering and Technology (MIET).

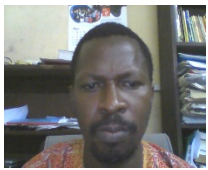
**Francis Okewole** was awarded the Bachelor of Technology (B. Tech.) Degree in Electrical Engineering, by the Ladoke Akintola University of Technology, Ogbomosho, Nigeria, in 2002, and obtained the degree of M. Sc. (also in Electrical Engineering) from the University of Lagos, in 2008. He is a Lecturer in the Department of Electrical and Electronics Engineering of the University of Lagos, and also a staff Ph.D. candidate, whose thesis on a hybrid MoM-PO analytical tool is ready for examination. Engr. Okewole has published a number of journal papers and conference proceedings papers in the area of antennas and radiowave propagation.

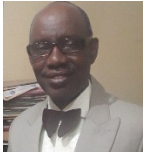
**Michael Adelabu** received the degree of M. Sc. in Electronics and Telecommunications from the Wroclaw Technical University in Poland in 1984, and Ph.D. (Electrical Engineering) from the University of Lagos in 2016. Dr. Adelabu is a Senior Lecturer in Electrical Engineering and has been with the University of Lagos since 1993. His current research interests extend to mobile communication networks and systems, radiowave propagation and spectrum management (efficient use). A fellow of the Nigerian Society of Engineers (FNSE), Dr. Adelabu is also a senior member of the IEEE.

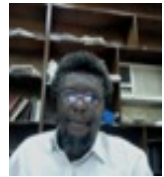
**Ike Mowete** obtained the degrees of B.Sc. (1980), M. Sc. (1983), and Ph.D. (1990) all in electrical engineering and from the University of Lagos, Akoka, Lagos, in Nigeria. His Ph.D. thesis was on a "quasi-static moment-method analysis of microstrip antennas". He has published quite a few papers on microstrip antennas, dielectric-costed thin-wire antennas, and shielding effectiveness of planar shields. His current research interests include thin-wire antenna structures, applications of the theory of characteristic modes to antenna analysis and design, and spectrum engineering issues. Professor Mowete, who teaches numerical methods, circuit theory and antennas and propagation at the University of Lagos is a member of the IEEE.